\documentclass[11pt]{article}
\usepackage{geometry}                
\usepackage{amsmath}
\usepackage{graphicx}
\usepackage{dcolumn}
\usepackage{bm}
\usepackage{wrapfig}
\usepackage{epsfig}
\usepackage{breqn}
\usepackage{amssymb}
\usepackage{subfigure}
\usepackage[usenames,dvipsnames]{pstricks}
\usepackage{epsfig}
\usepackage{pst-grad} 
\usepackage{pst-plot} 
 \usepackage[usenames,dvipsnames]{pstricks}
 \usepackage{epsfig}
 \usepackage{pst-grad} 
 \usepackage{pst-plot} 
 \usepackage{float}
 \usepackage{cite}

\usepackage[all]{xy}

\begin{document}

\title{Einstein's Field Equations as a Fold Bifurcation}
\author{Ikjyot Singh Kohli \\ isk@mathstat.yorku.ca \\ York University - Department of Mathematics and Statistics
\and Michael C. Haslam \\ mchaslam@mathstat.yorku.ca \\ York University - Department of Mathematics and Statistics}

\date{July 1, 2016}

\maketitle 

\begin{abstract}
It is shown that Einstein's field equations for \emph{all} perfect-fluid $k=0$ FLRW cosmologies have the same form as the topological normal form of a fold bifurcation. In particular, we assume that the cosmological constant is a bifurcation parameter, and as such, fold bifurcation behaviour is shown to occur in a neighbourhood of Minkowski spacetime in the phase space. We show that as this cosmological constant parameter is varied, an expanding and contracting de Sitter universe \emph{emerge} via this bifurcation. 
\end{abstract}

\section{Introduction}
The Einstein field equations of General Relativity are a highly nonlinear system of partial differential equations. In particular, for a four-dimensional spacetime, they consist of a coupled set of ten, nonlinear, hyperbolic partial differential equations. By examining Einstein's equations (in combination with Killing's equations) in the context of different types of matter configurations, one is able to form very detailed studies of the dynamics of cosmological models, which are largely governed by classical general relativity. It is then expected that many of the strong nonlinearities are present in these cosmological models as well \cite{Rugh:1994wk}. 

In fact, in considering spatially homogeneous cosmological models, the Einstein field equations themselves become a coupled system of nonlinear \emph{ordinary} differential equations \cite{ellismac}, and are thus ripe for study using various techniques from dynamical systems theory \cite{ellis}.

One such area of interest is in the possible applicability of bifurcation theory to these cosmological models. Bifurcations naturally arise in such models, as the majority of these cosmological models are dependent on at least one free parameter. As such, it is interesting to see how the local stability of fixed points of the Einstein field equations changes with respect to variations of these parameters. Numerous bifurcations have been discovered in the literature. The interested reader is asked to refer to \cite{ellis} and \cite{elliscosmo} and references therein for various examples. 

Further, with respect to bifurcation theory, we note that a very detailed study of some of the more ``exotic'' bifurcations has not been completed with respect to spatially homogeneous cosmological models. These types of bifurcations include fold bifurcations, Hopf bifurcations, cusp bifurcations, Bautin bifurcations, and Bogdanov-Takens bifurcations \cite{kuznet}. 
Examples of fold bifurcations in different cosmological models were found in \cite{Chavanis:2016dab, Szydlowski:2013sma, Abdelwahab:2007jp, vandenHoogen:1996pk}. 

In this paper, we will show that in fact, Einstein's field equations for spatially flat Friedmann-Lema\^{i}tre-Robertson-Walker (FLRW) universes have the same form as a topological normal form of a fold bifurcation, and that this behaviour is generic for \emph{all} possible perfect fluid matter configurations. Note that, throughout, we use units where $c=8\pi G = 1$. 

\section{The Dynamical Equations}
A cosmological model must be specified by a pseudo-Riemannian manifold, $\mathcal{M}$, a metric tensor, $g$, and a four-velocity field, $u$ that describes the matter flow in the model. The matter in the cosmological model is related to the spacetime geometry via the Einstein field equations:
\begin{equation}
R_{ab}  - \frac{1}{2}Rg_{ab} + \Lambda g_{ab} = T_{ab}.
\end{equation}
Following \cite{ellis, hervik}, one can then further decompose the covariant derivative of the four-velocity field as
\begin{equation}
u_{a;b} = \frac{1}{3}\theta h_{ab} + \sigma_{ab} + \omega_{ab} - \dot{u}_{a}u_{b},
\end{equation}
where $\theta$ is known as the expansion scalar, $h_{ab}$ is the projection tensor, $\sigma_{ab}$ is the kinematic shear tensor, $\omega_{ab}$ is the vorticity tensor, and $\dot{u}_{a}$ is the four-acceleration field. 

In this paper, we are concerned with $k=0$ FLRW spatially homogeneous and isotropic cosmological models, and as such, we have that $\sigma_{ab} = \omega_{ab} = \dot{u}_{a} = 0$. Further, this implies that the energy-momentum tensor must have the form of a perfect fluid:
\begin{equation}
T_{ab} = \mu u_{a} u_{b} + p h_{ab},
\end{equation}
where $\mu$ represents the energy density and $p$ the pressure. We make the further assumption that the matter in this model obeys a barotropic equation of state, so that we can write $p = w \mu$, where $-1 < w \leq 1$ is an equation of state parameter, and describes different types of matter configurations in the cosmological model under consideration. For example, $w = 1/3$ describes a radiation-dominated universe, while $w = 0$ describes a universe with pressure-less dust. 

The Einstein field equations then imply the energy-momentum conservation equation
\begin{equation}
\label{eq:enmom1}
\dot{\mu} + \theta\left[\mu(1+w)\right] = 0,
\end{equation}
Raychaudhuri equation
\begin{equation}
\label{eq:raych1}
\dot{\theta} + \frac{1}{3}\theta^2 + \frac{1}{2}\left[\mu(1 + 3 w)\right] - \Lambda = 0,
\end{equation}
and Friedmann equation
\begin{equation}
\label{eq:friedmann1}
\frac{1}{3}\theta^2 = \mu + \Lambda - \frac{1}{2} \mathcal{R},
\end{equation}
where $\mathcal{R}$ denotes the three-dimensional Ricci scalar (related to the spacetime foliation) and $\Lambda$ denotes the cosmological constant.

Since we are interested in $k=0$ FLRW models, one can in fact set $\mathcal{R} = 0$ in Eq. \eqref{eq:friedmann1}. We then can use Eq. \eqref{eq:friedmann1} in Eq. \eqref{eq:raych1} to get a single first-order ordinary differential equation that fully describes the dynamics of such models:
\begin{equation}
\label{eq:raych2}
\dot{\theta} = -\frac{1}{2} \left(w + 1\right) \left(\theta^2 - 3 \Lambda\right).
\end{equation}

\section{Fixed Points and The Fold Bifurcation}
With Eq. \eqref{eq:raych2} in hand, we will assume that $w$ is not a \emph{free} parameter, rather, it is fixed such that $-1 < w \leq 1$, and will denote it by $w_{0}$. We will also assume that the cosmological ``constant'', $\Lambda$ is a free parameter, where $\Lambda \in \mathbb{R}$. Therefore, Eq. \eqref{eq:raych2} describes the dynamics of spatially flat, FLRW perfect fluid models with an arbitrary cosmological constant. 

Eq. \eqref{eq:raych2} in this notation takes the form
\begin{equation}
\label{eq:raych2}
\dot{\theta} = -\frac{1}{2} \left(w_{0} + 1\right) \left(\theta^2 - 3 \Lambda\right).
\end{equation}

The fixed points of Eq. \eqref{eq:raych2} are found to be
\begin{equation}
\label{eq:fixed1}
\theta^{*} = \pm \sqrt{3 \Lambda},
\end{equation}
which describe expanding and contracting de Sitter universes respectively.

We will denote the right-hand-side of Eq. \eqref{eq:raych2} by $f(\theta,\Lambda)$. Then, motivated by Theorems 3.1 and 3.2 in \cite{kuznet}, from Eq. \eqref{eq:raych2} and the fixed points described in Eq. \eqref{eq:fixed1}, we see the following
\begin{enumerate}
	\item $f(\theta, \Lambda)$ is a polynomial in $\theta$, hence, is a smooth function.
	\item At $\Lambda = 0$, there exists a fixed point $\theta^* = 0$.
	\item Finally, we have that $\lambda = f_{,\theta}(0,0) = 0$, (where a comma subscript indicates a partial derivative).
	\item $f$ obeys the \emph{non-degeneracy} condition: 
		\begin{equation*}
			\frac{1}{2} \frac{\partial^2 f}{\partial \theta^2}|_{(\theta^*, \Lambda^*) = (0,0)} = \frac{1}{2} \left(-1 - w\right) \neq 0.
		\end{equation*}
	\item $f$ obeys the additional \emph{non-degeneracy} condition: 
		\begin{equation*}
		\frac{\partial f}{\partial \Lambda}|_{(\theta^*, \Lambda^*) = (0,0)} = \frac{3}{2} (w_{0} + 1) \neq 0,
		\end{equation*}
\end{enumerate}
where the last two inequalities follow from the fact that we are considering $-1 < w_{0} \leq 1$. 

Let us now also consider the following transformations. Namely,
\begin{equation}
\eta = \left| -\frac{1}{2} (w_{0} + 1) \right| \theta, \quad \beta =  \left| -\frac{1}{2} (w_{0} + 1) \right| \frac{3}{2}\Lambda (w_{0} + 1).
\end{equation}
Substituting these into Eq. \eqref{eq:raych2}, we obtain
\begin{equation}
\label{eq:normform2}
\dot{\eta} = \beta + s \eta^2.
\end{equation}
where
\begin{equation}
s = \frac{-\frac{1}{2}(w_{0} + 1)}{\left| -\frac{1}{2}(w_{0}+1)\right|} = \mbox{sign} \left[\frac{1}{2}(-1 - w_{0})\right].
\end{equation}
Note that $s = -1$ for $-1 < w_{0} \leq 1$, that is, for cosmological models with ordinary matter. Further $s = 1$ for $w_{0} <-1$, which describes a cosmological model with phantom energy \cite{hervik, nojiri3}. 

Indeed, Eq. \eqref{eq:normform2} is precisely the topological normal form for a fold bifurcation. Further, the five conditions described after Eq. \eqref{eq:fixed1} show that the fold bifurcation behaviour is in fact a \emph{generic} property of the Einstein field equations with respect to the cosmological models considered in this work.

\section{Implications of the Fold Bifurcation}
\begin{figure}[h]
\centering
\includegraphics[scale=0.55]{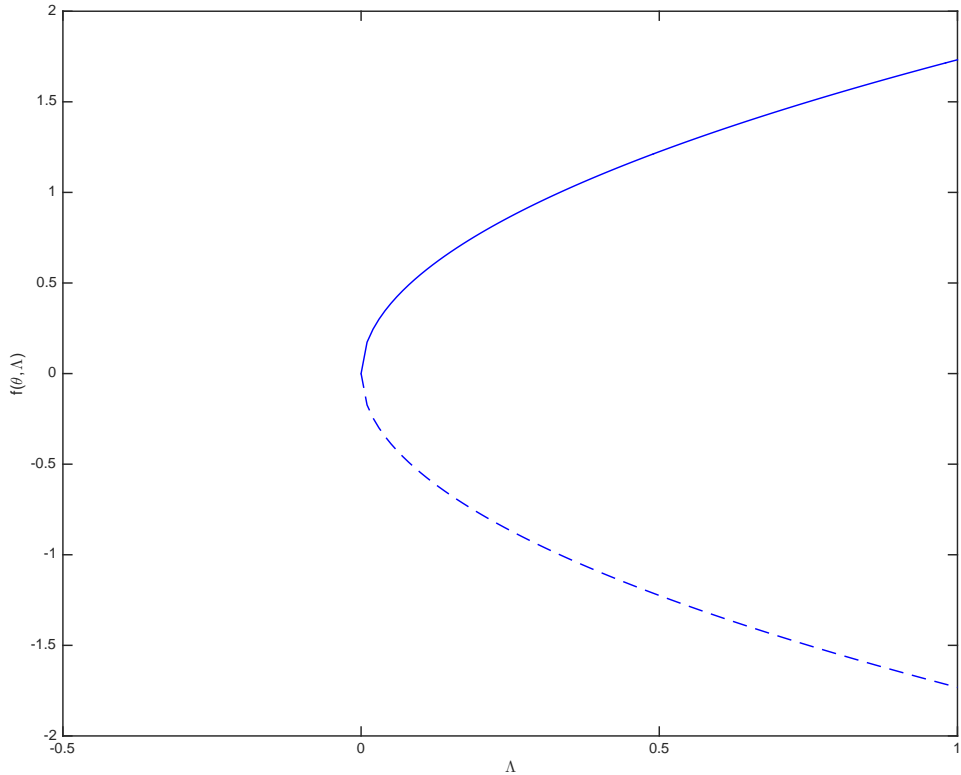}
\caption{A plot of the equilibrium manifold $f(\theta, \Lambda) = 0$ that clearly shows the fold bifurcation behaviour for varying values of $\Lambda$.}
\label{fig:fig1}
\end{figure}

The existence of a fold bifurcation as described above has some interesting physical effects. First, one can observe this bifurcation behaviour by plotting the equilibrium manifold, which is given by $f(\theta,\Lambda) = 0$ and shown in Figure \ref{fig:fig1}.

One sees from Figure \ref{fig:fig1} that there are no fixed points for $\Lambda < 0$, one fixed point for $\Lambda = 0$ which is the Minkowski spacetime, and two fixed points for $\Lambda > 0$, which are the expanding and contracting de Sitter universes described earlier. One sees that as $\Lambda$ passes through $\Lambda = 0$, two equilibrium points are \emph{created}. Note that the Jacobian is given by
\begin{equation}
f'(\theta) = -(1+w_0)\theta,
\end{equation}
At the de Sitter points, we have that
\begin{equation}
f'(\pm\sqrt{3 \Lambda}) = \mp(1+w_0) \sqrt{3 \Lambda}. 
\end{equation}
Therefore, the expanding de Sitter universe is an unstable node if $w_0 < -1$ and is a stable node if $-1 < w_0 \leq 1$. Further, the contracting de Sitter universe is an unstable node for $-1 < w_0 \leq 1$, while it is a stable node if $w_0 < -1$. 

We see that even a slight/perturbed variation of the cosmological constant has drastic effects on the overall dynamics. In some strong sense, we see that the fold bifurcation allows the de Sitter spacetime, and hence the standard inflationary scenario \cite{elliscosmo} to emerge from a Minkowski spacetime. This could have strong connections to some of the detailed studies completed on emergent universes \cite{Ellis:2002we, 2003magr.workE..13M, 2004CQGra..21..233E, 2005gr.qc.....5103M,2005PhRvD..71l3512M,2006CQGra..23.6927M,2007PhRvL..98z1301E,2008mgm..conf.1873M,2009CQGra..26g5017B,2010JCAP...06..026D,2011IJMPD..20.2767G,2011IJTP...50...80C,2011MNRAS.413..686P,2011IJTP...50.2708M,2012Ap&SS.339..101S,2012PhRvD..86d3518L,2012PhLB..718..248C,2013IJMPD..2230018G,2013PhRvD..88j3504A,2014JCAP...01..048Z,2014PhLB..732...81C,2015CQGra..32k5001P,2015PhRvD..92l4067B,2016IJTP..tmp..172G}, and warrants a more detailed investigation.

\section{Conclusions}
In this paper, we showed that Einstein's field equations for \emph{all} perfect-fluid $k=0$ FLRW cosmologies have the same form as the topological normal form of a fold bifurcation. In particular, in assuming that the cosmological constant was a bifurcation parameter, we showed that the fold bifurcation behaviour is shown to occur in a neighbourhood of Minkowski spacetime point in the phase space. We then showed that as this cosmological constant parameter is varied, an expanding and contracting de Sitter universe \emph{emerge} via this bifurcation.

\section{Acknowledgements}
This research was partially supported by a grant given to MCH from the Natural Sciences and Engineering Research Council of Canada. 

\newpage 
\bibliographystyle{ieeetr} 
\bibliography{sources}

\end{document}